# Nearly Isotropic Quantum-Critical Transport in Single-Crystal $CeNiC_2$

Hanming Ma[1,2,3], Jun Gouchi[3], Dilip Bhoi[3], Toru Shigeoka[4], Bosen Wang[1,2], J.-G. Cheng[1,2], and Yoshiya Uwatoko[3,5]

*[1]Beijing National Laboratory for Condensed Matter Physics and Institute of Physics, Chinese Academy of Sciences, Beijing 100190, China*

*[2]School of Physical Sciences, University of Chinese Academy of Sciences, Beijing 100049, China*

*[3]Institute for Solid State Physics, The University of Tokyo, Kashiwa, Chiba 277-8581, Japan*

*[4]Graduate School of Science and Technology, Yamaguchi University, Yamaguchi 753-8512, Japan*

*[5]Faculty of Science and Engineering, Tokyo City University, Setagaya, Tokyo 158-8557, Japan*

Correspondence: yuwatoko@tcu.ac.jp (YU); bswang@iphy.ac.cn (BSW); jgcheng@iphy.ac.cn (JGC)

**Abstract.** Pressure-induced superconductivity and *T*-linear resistivity have been reported in polycrystalline $CeNiC_2$, but orientational averaging has left the directional character of the critical scattering unresolved. We report pressure-dependent resistivity of high-quality single crystals for current along each crystallographic axis. These crystals have substantially lower residual resistivity and a slightly higher maximum onset $T_c$ than the polycrystalline sample, placing superconductivity in a cleaner transport regime. Near $P_c \approx 9.5$–$10$ GPa, the normal-state resistivity becomes nearly T-linear along every axis, the fitted residual resistivity is strongly enhanced, and superconductivity forms a narrow dome. For *I*//*b*, the *T*-linear normal state remains nearly unchanged in magnetic fields up to 9 T applied along *a* and *c*; the upper critical field is large and only moderately anisotropic. The common evolution along all three axes establishes a nearly isotropic quantum-critical transport regime, inconsistent with a simple low-dimensional spin-fluctuation picture and implicates valence fluctuations as the leading source of critical scattering associated with the superconducting dome.

## I. INTRODUCTION

Quantum criticality in a heavy-fermion metal arises when a nonthermal control parameter continuously suppresses an ordered ground state to zero temperature. Its central issue is not simply the disappearance of magnetic order, but the fate of the Kondo-coupled *f* electrons at that boundary. In the itinerant spin-density-wave description, Kondo screening remains intact and the quantum phase transition is an instability of the heavy Fermi surface; critical magnetic fluctuations are concentrated near the ordering wave vector and scatter selected regions of the Fermi surface. In a Kondo-destruction scenario, by contrast, the Kondo entanglement itself becomes critical and the Fermi surface reconstructs together with the loss of magnetic order, producing a more local form of electronic criticality [1-3]. A pressure-tuned valence instability provides a further nearly local channel in the charge sector [4-7]. Superconductivity may then emerge as an instability promoted by these critical fluctuations; it is a possible response to the critical normal state rather than an alternative origin of quantum criticality. The central experimental task is therefore to identify which electronic degree of freedom becomes critical and what momentum-space structure its scattering possesses.

$CeNiC_2$ provides a concrete setting in which to make this distinction. At ambient pressure it undergoes incommensurate and commensurate antiferromagnetic transitions near 20 and 10 K, followed by a low-temperature canted ferromagnetic component [8-10]. Under pressure, the magnetic order is first enhanced and then suppressed near 11 GPa, where a narrow superconducting dome reaches $T_c \approx 3.5$-$3.8$ K, $\mu_0 H_{c2}(0)$ greatly exceeds the weak-coupling Pauli scale, and the polycrystalline normal-state resistivity becomes nearly T linear [11]. Single-crystal x-ray diffraction shows that the crystallographic symmetry persists to 18.6 GPa, while the

first- and second-nearest Ce-Ce (and Ni-Ni) directions interchange near 7 GPa, close to the maximum of the magnetic-ordering temperature [12]. Pressure therefore reorganizes the bond network while tuning the magnetic ground state, making $CeNiC_2$ a well-defined case in which to determine whether the critical transport is governed by wave-vector-selective magnetic scattering or by a more local electronic channel.

Electrical resistivity is a sensitive probe because its temperature exponent and directional dependence encode the spatial structure of the dominant scattering. For a conventional three-dimensional antiferromagnetic quantum critical point, self-consistent renormalization theory predicts $\Delta\rho \propto T^{3/2}$; T-linear resistivity is instead expected for effectively lower-dimensional or nearly local scattering [4–7, 13, 14]. Wave-vector-selective fluctuations should therefore retain a directional imprint of the magnetic and electronic structure, whereas a nearly local channel should produce a common evolution for different current directions. The central question in $CeNiC_2$ is whether the T-linear resistivity found in the polycrystal is direction selective or common to all three crystallographic axes. Our single-crystal data show that the exponent $n$ in $\Delta\rho \propto T^n$ approaches 1 and that the fitted $\rho_0$ is strongly enhanced at the same pressure scale for every current direction. This convergence reveals a nearly axis-uniform critical channel and identifies valence fluctuations as a leading microscopic source of the critical scattering.

## II. EXPERIMENTAL METHODS

We grew single crystals of $CeNiC_2$ by the Czochralski pulling method and verified their phase purity and crystallographic orientations by single-crystal x-ray diffraction. We measured magnetic susceptibility and ambient-pressure resistivity using Quantum Design MPMS and PPMS platforms. Electrical resistivity was measured by a four-probe current-reversal method with Au wires attached by silver paste. We prepared oriented specimens with current along the *a*, *b*, and *c* axes. Zero-field measurements from 2 to 300 K were performed in a constant-loading cubic-anvil apparatus at ISSP (UTokyo), to 13 GPa for $I \parallel a$ and to 12 GPa for $I \parallel b$ and $c$ [15]. During each temperature sweep, the applied load was actively regulated to maintain constant pressure. Pressure was determined from calibrated resistance-load relations for Sn, Bi, and Pb at room, liquid-nitrogen, and liquid-helium temperatures [15, 16]. A 1:1 volume mixture of Fluorinert FC70 and FC77 was used as the pressure-transmitting medium.

Field-dependent measurements near the superconducting dome were performed in a clamp-type palm cubic-anvil cell [17, 18] mounted in a dilution refrigerator. With $I \parallel b$, we measured $\rho(T)$ down to 20 mK in fields up to 9 T for $H \parallel a$ at 10.4 GPa and $H \parallel c$ at 10.5 GPa; the two field orientations were measured in separate loadings. The three current directions were also measured in independent pressure runs. Comparisons among directions used the nearest pressure points together with the common evolution across each pressure series. For the analysis, the resistivity data were averaged in 0.5-K intervals. Power-law fits started at 5 K, above the superconducting transition, and excluded magnetic-transition regions. We obtained $\rho_0$ and $n$ from $\rho(T) = \rho_0 + AT^n$ and cross-checked $n$ using $n = d \ln[\rho(T) - \rho_0]/d \ln T$.

## III. RESULTS AND DISCUSSION

### A. Ambient-pressure anisotropy

Figure 1 summarizes the ambient-pressure response of magnetic susceptibility and electrical resistivity. The susceptibility $\chi_i(T)$ ($i$ = *a*, *b*, *c*) shows the same sequence of anomalies in every direction. Peaks near 19 and 11 K identify the incommensurate and commensurate antiferromagnetic transitions. Curie-Weiss fits above the ordered state give effective moments of approximately 2.48, 2.54, and 2.49 $\mu_B$/Ce for fields along *a*, *b*, and *c* axes, respectively, close to the $Ce^{3+}$ free-ion value. The strongly anisotropic Curie-Weiss temperatures, about -2, -55, and -5 K, reflect the crystalline-electric-field anisotropy. Magnetization remains nearly linear to 5 T; only the *b*-axis curve shows a weak low-field ferromagnetic component, in agreement with the reported canted state [8, 9].

Likewise, the resistivity follows a common temperature dependence but has a direction-dependent magnitude, $\rho_b > \rho_c > \rho_a$. The incommensurate transition produces a pronounced drop near 20 K, and the commensurate transition gives a weaker feature near 10 K. Residual-resistivity ratios are approximately 23, 50, and 39 for current along *a*, *b*, and *c*, respectively. The corresponding residual resistivities, 2.46, 5.02, and 4.42 μΩ cm, are all lower than the 9.85 μΩ cm reported for the polycrystalline specimen [11]. Together with the sharper transition signatures, these values demonstrate that the single crystals provide a cleaner platform for resolving the directional structure of the pressure-tuned scattering.

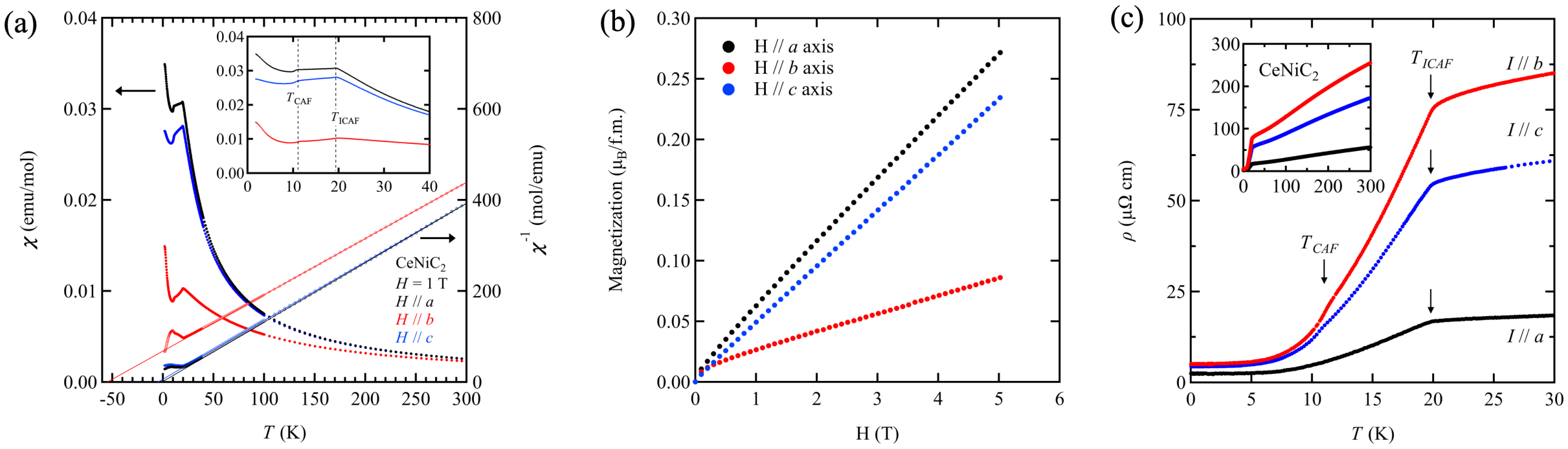

**FIG. 1.** Ambient-pressure anisotropy of $CeNiC_2$. (a) Magnetic susceptibility and inverse susceptibility along the three axes; the inset expands the magnetic transitions. (b) Magnetization at 1.8 K. (c) Low-temperature resistivity, with the 0–300 K range in the inset. All axes share a common temperature evolution with direction-dependent magnitudes.

## B. Pressure evolution

Figure 2 shows $\rho_i(T)$ ($i = a$, $b$, and $c$) at pressures up to 12–13 GPa. The room-temperature resistivity rises with pressure for all three current directions. At lower temperatures, the incommensurate ordering temperature $T_{ICAF}$ first increases from about 19 K to a maximum near 34 K at approximately 7 GPa and then decreases, becoming unresolved near 10 GPa. At the same time, a broad resistivity maximum associated with the Kondo/coherence scale $T_K$ appears between 4 and 6 GPa and shifts rapidly to higher temperature above about 8 GPa; beyond 10 GPa, it lies above our 300 K measurement range. The opposing pressure dependences of $T_{ICAF}$ and $T_K$ reveal a crossover from magnetic-order-dominated behavior to increasing Kondo hybridization. Their common change near 7 GPa coincides with the reported interchange of the first- and second-nearest Ce-Ce directions [12]. This coincidence links the bond-network reorganization to the pressure evolution of hybridization and magnetism within the unchanged crystallographic symmetry.

Superconductivity emerges as $T_{ICAF}$ collapses. For $I \parallel a$, a resistive superconducting onset appears near 9 GPa and then reaches about 3.8 K at 9.7 GPa, followed by zero resistance over the narrow 9.5–10 GPa interval. For $I \parallel b$, the resistive onset $T_c$ reaches a maximum of about 3.84 K near 9.6 GPa, with zero resistance extending to roughly 10.4 GPa. For $I \parallel c$, the resistive onset $T_c$ reaches about 3.9 K near 9.5 GPa. Thus, the three current directions define the same narrow superconducting dome centered on the collapse of magnetic order. The maximum resistive onset $T_c$ is slightly higher than the approximately 3.5–3.8 K range reported for the polycrystal [11], while the residual resistivity is substantially lower in the single crystal. The single-crystal results therefore reveal superconductivity in a cleaner transport regime and underscore the sensitivity of the pairing state to disorder.

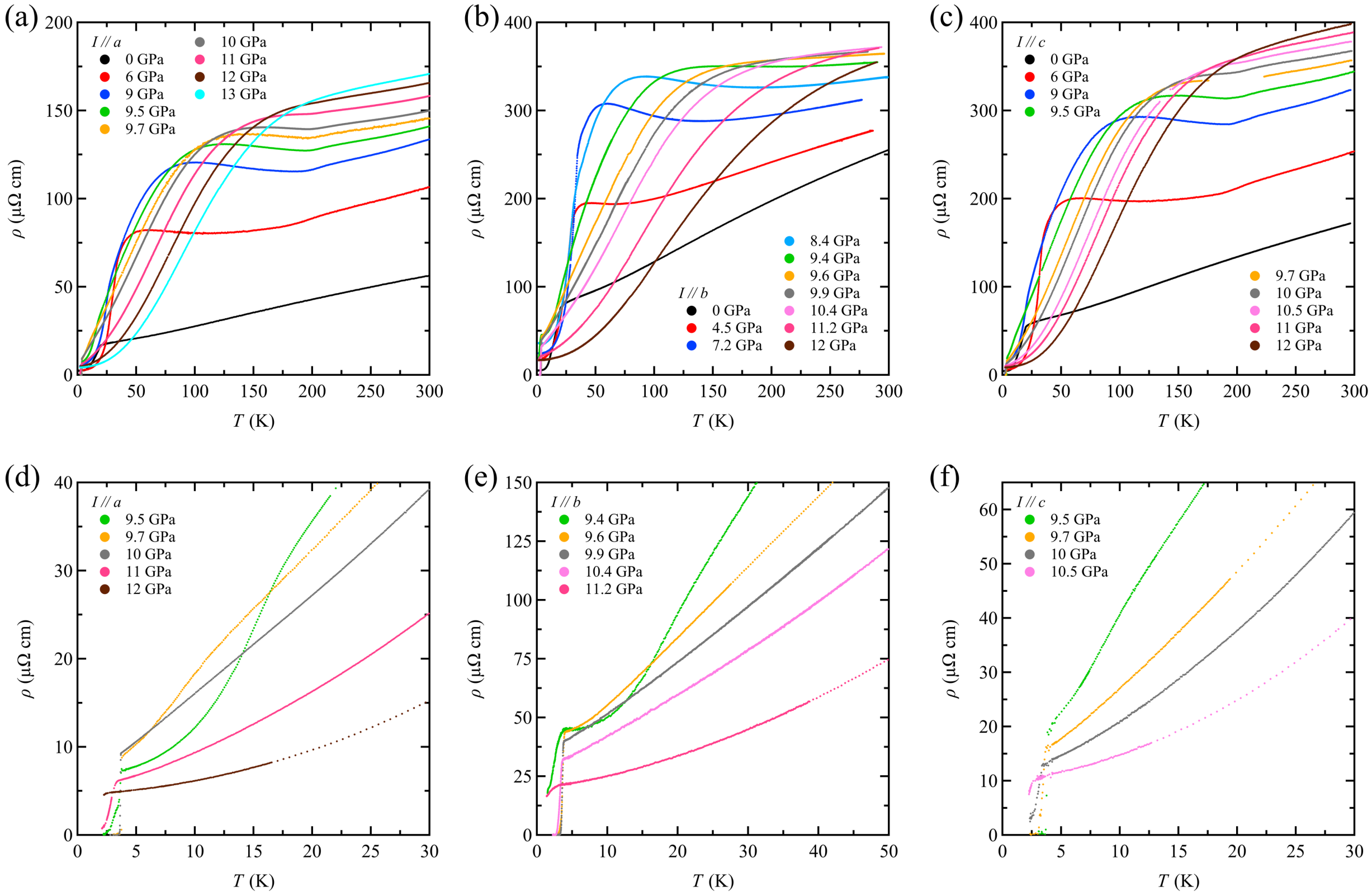

**FIG. 2.** Temperature-dependent resistivity under pressure for (a, d) *I*//*a*, (b, e) *I*//*b*, and (c, f) *I*//*c*. The upper row shows the full temperature range; the lower row expands the superconducting and magnetic transitions. All three orientations show the same sequence: enhancement and suppression of magnetic order, growth of the Kondo/coherence scale, and a narrow superconducting dome near 9.5-10 GPa.

## C. Quantum-critical regime

The central result of the three-axis comparison is the common pressure evolution observed for all three current directions. Figure 3 (a) and 3(b) show the field dependence of the normal state with external fields applied along the *a* and *c* axis, respectively. Applying a magnetic field suppresses the superconducting transition and recovers the low-temperature normal state. With *I*//*b*, fields up to 9 T shift the superconducting transition monotonically to lower temperature while leaving the *T*-linear temperature dependence of the normal-state resistivity nearly unchanged. Figure 3 (c) shows the fitting results of $H_{c2}(T)$ using the Werthamer-Helfand-Hohenberg form [19] give $\mu_0 H_{c2}^a(0) \approx 19.8$ T for *H*//*a* at 10.4 GPa and $\mu_0 H_{c2}^c(0) \approx 15.0$ T for *H*//*c* at 10.5 GPa. Both values exceed the weak-coupling Pauli field, and their ratio of about 1.3 indicates moderate anisotropy of the Fermi surface. The field-resilient *T*-linear resistivity and large $H_{c2}$ link the quantum-critical normal state directly to unconventional superconductivity. Figure 3(d) shows that the fitted $\rho_0(P)$ develops a pronounced maximum between 9.5–10 GPa; its absolute value remains largest for *I*//*b*, while its pressure evolution is common to all axes. Also, direct power-law fits and the logarithmic-derivative analysis of $\rho$(T) give the same evolution of *n* for all three axes. The simultaneous occurrence of *T*-linear transport and enhanced $\rho_0$ is characteristic of valence-critical transport and implicates valence fluctuations as a leading microscopic source of critical scattering [4–6].

Compared with the earlier polycrystalline transport, the single-crystal measurements provide two key results. First, the three directional exponents and residual-resistivity enhancements converge on the same critical pressure range despite different absolute resistivities along different crystallographic directions. Second, independent of crystal axis, applied pressure reproduces the same sequence of magnetic order, growth of the Kondo/coherence scale, superconductivity, and $T$-linear normal-state transport. Together, these results constrain the critical mechanism responsible for superconductivity. In addition, the lower residual resistivity, clear zero-resistance transitions, and slightly higher maximum onset $T_c$ show that the superconducting dome persists in a cleaner single-crystal transport regime.

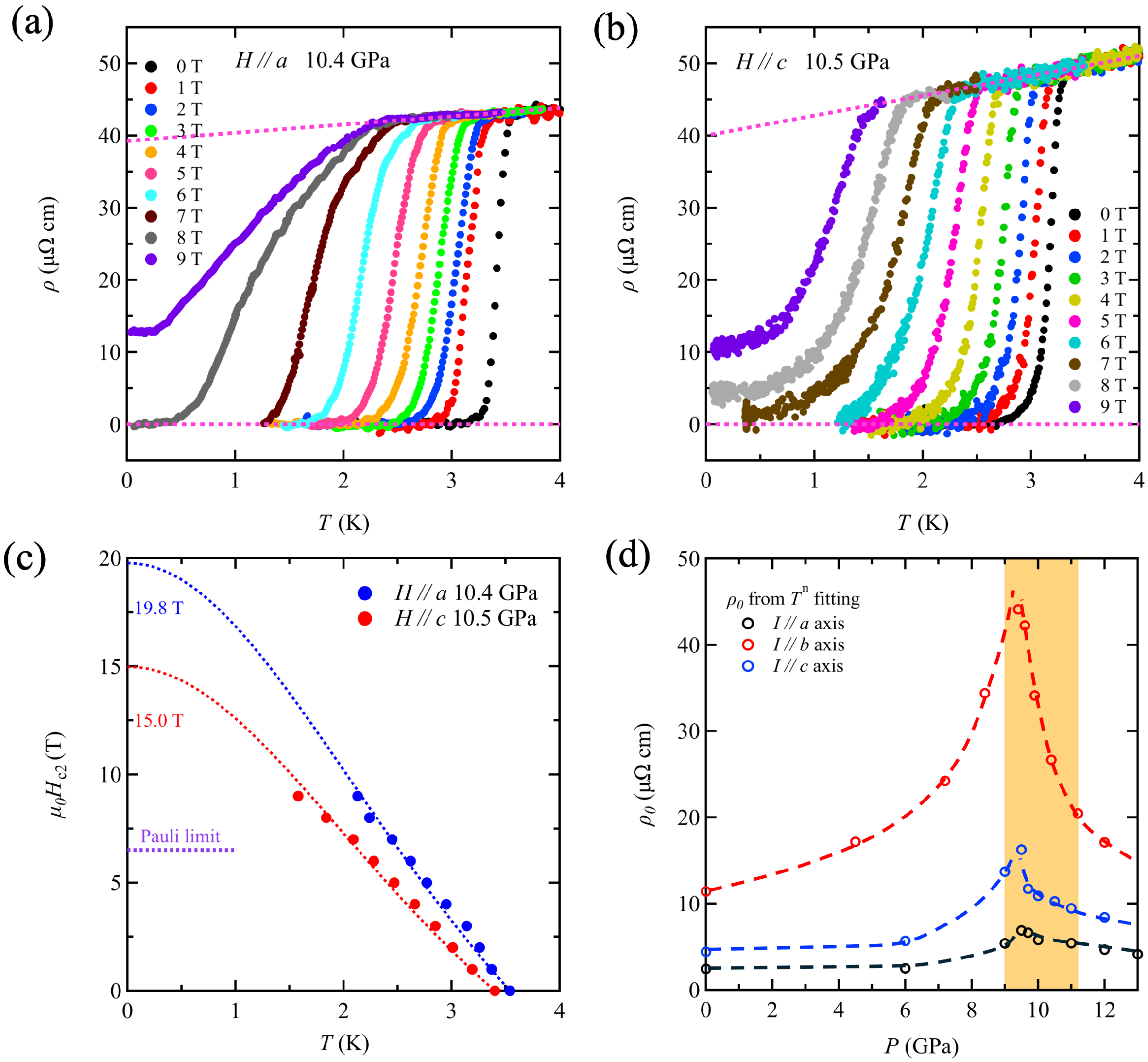

**FIG. 3.** Field and critical-scattering diagnostics near the superconducting dome. Field-dependent resistivity for (a) $H//a$ at 10.4 GPa and (b) $H//c$ at 10.5 GPa, with $I//b$. (c) Upper critical fields with WHH extrapolations and the weak-coupling Pauli scale. (d) Fitted residual resistivity $\rho_0(P)$ for the three current directions, showing a common enhancement in the critical pressure range.

Figure 4 summarizes the pressure-temperature phase diagrams and $n(T,P)$. The central result is that measurements for the three current directions reveal the same topology: (i) $T_{ICAF}$ forms a broad maximum near 7 GPa and becomes unresolved near 9.5–10 GPa; (ii) $T_K$ rises steeply as magnetic order is lost; and (iii) the superconducting dome overlaps the region with $n \approx 1$. A conventional three-dimensional antiferromagnetic critical point gives $n = 3/2$, whereas effectively two-dimensional antiferromagnetic scattering produces T-linear transport primarily along the fluctuation-dominated directions [13, 14, 20, 21]. $CeNiC_2$ instead exhibits T-linear transport along all three crystallographic axes, revealing a nearly local critical-scattering channel beyond the simplest low-dimensional spin-fluctuation descriptions.

Valence fluctuations provide a common explanation for the two defining transport observations: T-linear resistivity along every axis and a pronounced enhancement of $\rho_0$ [4–7]. In $CeNiC_2$, $T_K$ and $T_{ICAF}$ change together near the pressure where the Ce-Ce bond network is reorganized, providing a microscopic link between pressure, hybridization, and valence-sensitive scattering [12]. Taken together, these results implicate valence fluctuations as a leading microscopic source of the nearly isotropic critical scattering and as a natural candidate for the pairing interaction. Pressure-dependent Ce L-edge x-ray absorption or resonant x-ray spectroscopy across 7–11 GPa can directly test this microscopic picture.

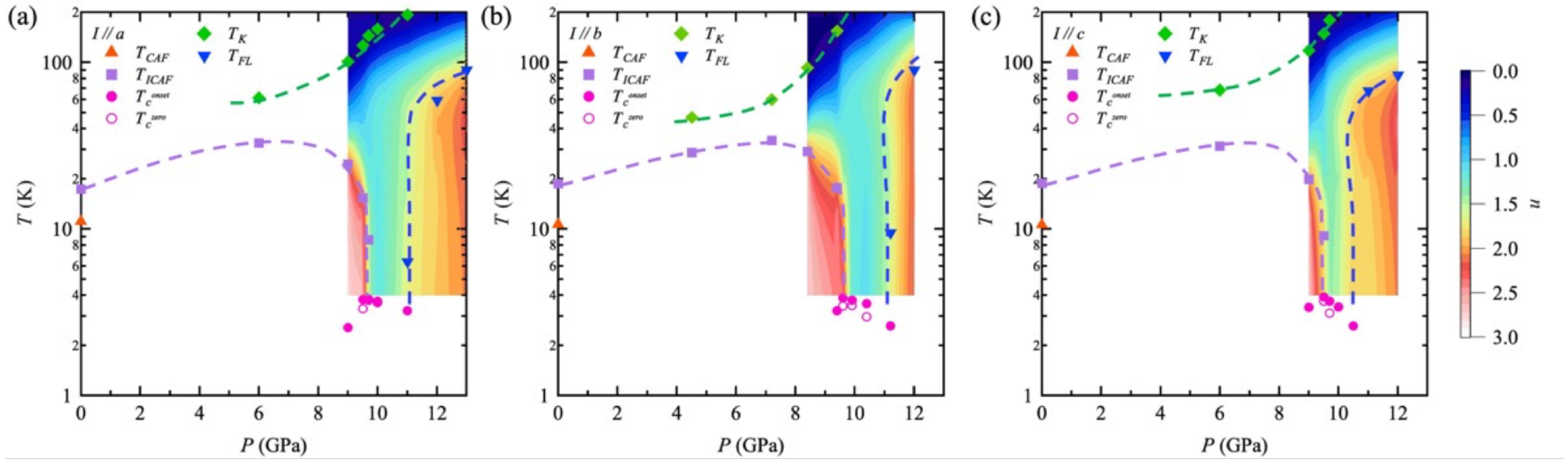

**FIG. 4.** Pressure-temperature phase diagrams for (a) *I*//*a*, (b) *I*//*b*, and (c) *I*//*c*. Symbols mark magnetic, superconducting, Kondo/coherence, and Fermi-liquid crossover scales. The color maps show $n(T,P)$ obtained from the logarithmic-derivative analysis after 0.5-K averaging. The common topology across the independent pressure series is the central directional result.

## IV. CONCLUSION

Our direction-resolved transport measurements using high quality $CeNiC_2$ single-crystal reveal the same critical evolution along all three crystallographic axes in $CeNiC_2$ under pressure. Magnetic order is first enhanced and then suppressed, the Kondo/coherence scale rises rapidly, and a narrow superconducting dome overlaps the pressure range in which $n$ approaches 1 and $\rho_0$ is enhanced. With *I*//*b*, the T-linear normal-state resistivity remain nearly unchanged in fields up to 9 T, while $H_{c2}$ is large and only moderately anisotropic. The common three-axis evolution establishes a nearly isotropic quantum-critical transport regime beyond simple low-dimensional spin-fluctuation descriptions. The lower residual resistivity and slightly higher maximum resistive onset $T_c$ reveal the superconducting dome in a cleaner single-crystal transport regime and underscore its sensitivity to disorder. Together with the Ce-Ce bond-network reorganization, the transport results implicate valence fluctuations as a leading microscopic source of the pressure-tuned criticality and as a natural candidate for the pairing interaction in $CeNiC_2$.

## ACKNOWLEDGMENTS

We thank M. Matsuda, F. Ye, and K. Matsubayashi for helpful discussions. This work was supported by JSPS KAKENHI Grant Numbers JP19H00648 and JP25K00951. Part of this work was carried out at the Synergetic Extreme Condition User Facility (SECUF, https://cstr.cn/31123.02.SECUF). This work was also supported in part by the CAS President's International Fellowship Initiative (Grant No. 2024PG0003), and the International Young Scientist Fellowship of the Institute of Physics, Chinese Academy of Sciences (Grant No. 202604).